\documentclass[10pt,aps,prl,reprint,groupedaddress,showkeys,superscriptaddress,amsmath,amssymb,onecolumn]{revtex4-2}

\pdfoutput=1

\usepackage[utf8]{inputenc}
\usepackage{amssymb,amsmath}
\usepackage{graphicx}
\usepackage{booktabs} 
\usepackage{csquotes} 
\usepackage{commath}
\usepackage[american]{babel}
\usepackage{url}

\usepackage{todonotes}
\setuptodonotes{inline}

\usepackage{color}
\definecolor{red}{rgb}{1,0,0}
\definecolor{blue}{rgb}{0,0,1}
\definecolor{green}{rgb}{0,0.5,0}

\usepackage{hyperref}
\hypersetup{
    colorlinks=true,       
    linkcolor=red,          
    citecolor=green,        
    filecolor=magenta,      
    urlcolor=cyan           
}

\begin{document}
\title{RidePy: A fast and modular framework for simulating ridepooling systems}

\author{Felix Jung} 
\email{felix.jung@tu-dresden.de}
\affiliation{Chair of Network Dynamics, Institute of Theoretical Physics and Center for Advancing Electronics Dresden (cfaed), TUD Dresden University of Technology, 01062 Dresden, Germany}

\author{Debsankha Manik}
\affiliation{Chair of Network Dynamics, Institute of Theoretical Physics and Center for Advancing Electronics Dresden (cfaed), TUD Dresden University of Technology, 01062 Dresden, Germany}

\date{\today}

\maketitle

\section{Summary}\label{summary}

RidePy enables fast computer simulations of on-demand mobility modes
such as ridehailing or ridepooling. It strongly focuses on modeling the
mobility service itself, rather than its customers or the environment.
Through a combination of Python \cite{vanrossum1995}, Cython \cite{cython}
and C++ \cite{stroustrup2000}, it offers
ease of use at high performance. Its modular design makes customization
easy, while the included modules allow for a quick start.

\section{Statement of need}\label{statement-of-need}

An accelerating climate change and congested cities both call for an
urgent change in the way we move \cite{winkler2023}. To reduce carbon dioxide emissions as well as the number of
vehicles on the road, digitally managed on-demand mobility services such
as ridehailing and ridepooling are explored in research
\cite{engelhardt2019, santi2014} and on the road.
Unfortunately, physically experimenting with such services for research
purposes is extremely cost- and time-intensive. However, the operational
properties of such systems are largely predefined in terms of the
scheduling backend that manages them. This makes it possible to replace
physical experiments with computer simulations, substituting virtual
vehicles for actual ones and modeling the incoming mobility demand by
sampling either historic requests or synthetic distributions. Another
advantage of simulations is that the degree to which they represent
reality may be freely adjusted. This makes it possible to both answer
concrete operational questions \cite{deruijter2023, henao2019a, lotze2022b, ruch2020, wilkes2021a, zwick2021a, zwick2022c} and investigate idealized system
behavior, gaining deeper insights into the general properties of
on-demand mobility systems \cite{herminghaus2019a, manik2020a, molkenthin2020, muehle2023, tachet2017a, zech2022}.

In this context, a simulation framework should appropriately allow for
vastly different system sizes and degrees of realism. The system size
incorporates the number of simulated vehicles as well as the extent of
the space they operate on: A small system may consist of a single
vehicle serving a network of just two nodes, while an example of a large
system could be a fleet of several thousand vehicles operating on the
street network of a large city. The degree of realism may be varied, for
example, by sampling requests from either a uniform distribution or
recorded mobility demand, or by operating on a continuous Euclidean
plane versus a realistic city street network. Another option is to
adjust the constraints imposed, such as the time windows assigned to
stops or the vehicles' seat capacities.

Finally, an on-demand mobility simulation framework should be fast, easy
to use and adaptable to various applications.

A number of open-source simulation software projects are already being
used to investigate on-demand mobility services. Some of them focus on
microscopic modeling in realistic settings, through which concrete
predictions for service operation are enabled, guiding urban planning.
Prominent examples are MATSim \cite{horni2016}, which performs agent-based simulations of individual
inhabitants, and Eclipse SUMO \cite{lopez2018}, a microscopic traffic simulator. Both rely on additional
packages to model on-demand mobility, such as AMODEUS
\cite{ruch2018} for MATSim and Jade
\cite{behrisch2014} for SUMO.

FleetPy \cite{engelhardt2022}, a
recently released on-demand mobility simulation, is primarily aimed at
realistic modeling of the interactions between operators and users,
specifically incorporating multiple operators. While its technical
approach is similar to ours, integrating Python with fast Cython and C++
extensions, the project is predominantly focused on applied simulations,
although its framework architecture promises to allow for adjustment of
the model detail level.

Perhaps the most idealized approach is taken by the Julia
\cite{bezanson2017} package
\texttt{RidePooling.jl} \cite{muehle2022} which
was developed in support of a recent scientific contribution
\cite{muehle2023}.

A very different yet interesting route is taken by MaaSSim
\cite{kucharski2022a}, which models
on-demand mobility in the realm of two-sided mobility platforms such as
Uber \cite{uber} and Lyft
\cite{lyft}.

RidePy extends this landscape by providing a universal and fast
ridepooling simulation framework that is highly customizable while still
being easy to use. It is focused on modeling the behavior of a vehicle
fleet while covering a broad scope in terms of system size and degree of
realism.

\section{Philosophy and usage}\label{philosophy-and-usage}

RidePy simulates flexible mobility services based on \emph{requests},
\emph{dispatchers} and \emph{vehicles}. The vehicles continuously move
along routes defined by scheduled \emph{stops}. At each stop, passengers
are picked up or dropped off, leading to a change in seat occupancy
aboard the vehicle. A \texttt{RequestGenerator} supplies requests for
mobility that are submitted to the simulated service, consisting of
origin and destination locations and optional constraints. A
\texttt{dispatcher} processes these incoming requests. If a request
cannot be fulfilled given the constraints (e.g., time windows, seat
capacity), it is rejected upon submission. Otherwise, pick-up and
drop-off stops are scheduled with a vehicle, respectively.

All individual components of the simulation framework may be customized
or replaced. This includes \texttt{RequestGenerator}s,
\texttt{dispatcher}s, and the \texttt{TransportSpace} which the system
operates on. Examples for \texttt{TransportSpace}s include the
continuous Euclidean plane and arbitrary weighted graphs (e.g., street
networks). Several components of RidePy are implemented in both pure
Python and Cython/C++. While their pure Python versions are easier to
understand, debug and modify, the Cython/C++ versions make large-scale
simulations tractable.

Running a RidePy simulation yields a sequence of \texttt{Event}s. The
included analytics code consumes these events and returns two extensive
Pandas \cite{wesmckinney2010}
\texttt{DataFrame}s: \texttt{stops} and \texttt{requests}.
\texttt{stops} contains all stops that have been visited by each
vehicle, along with additional information such as the vehicles'
passenger occupancy. \texttt{requests} similarly contains all requests
that have entered the system, enriched with secondary information such
as the time riders spent on the vehicle.

Additional included tooling allows for the setup, parallel execution,
and analysis of simulations at different parameters (parameter scans).
This includes the serialization of all simulation data in JSON format
\cite{json}.

To ensure valid behavior, RidePy incorporates an extensive automated
test suite \cite{pytest}.

\section{Availability}\label{availability}

RidePy is available from PyPI \cite{ridepy-pypi}. The source code is hosted on GitHub
\cite{ridepy-github}.
Extensive documentation can be found on the project's webpage
\cite{ridepy-doc}.

\section{Acknowledgements}\label{acknowledgements}

We kindly thank Philip Marszal, Matthias Dahlmanns, Knut Heidemann,
Malte Schröder, and Marc Timme for their input and advice.

This project was partially supported by the Bundesministerium für Bildung und Forschung (BMBF, German Federal Ministry of Education and Research) under grant No.~16ICR01 and by the Bundesministerium für Digitales und Verkehr (BMDV, German Federal Ministry for Digital and Transport) as part of the innovation initiative mFund under grant No.~19F1155A.

\section{Competing interests}

Debsankha Manik was employed at MOIA GmbH when this research was conducted. MOIA GmbH
neither sponsored nor endorses his research.

\end{document}